\documentclass[aps,prc,twocolumn,amsmath,superscriptaddress,floatfix,nofootinbib]{revtex4-2}

\usepackage{setspace,ulem, epsfig,amssymb,amsfonts,amsmath,mathtools,bm,color,xcolor,graphicx,braket,adjustbox,esint,upgreek,verbatim,ctable}

\usepackage[colorlinks, linkcolor=red,anchorcolor=green,citecolor=blue]{hyperref}

\newcommand{\tred}{\textcolor{black}}

\begin{document}
\title{Triple-charmed Hadron from Coalescence in Relativistic Heavy-Ion Collisions}

\author{Tianyang Li}
\affiliation{Department of Physics, Tianjin University, Tianjin 300350, China}

\author{Jiamin Liu}
\email{Co-corresponding author: liujiamin@tju.edu.cn}
\affiliation{Department of Physics, Tianjin University, Tianjin 300350, China}

\author{Shiqi Zheng}
\email{Co-corresponding author: sz3357@columbia.edu}
\affiliation{Department of Civil Engineering and Engineering Mechanics, Columbia University, New York, NY 10027, USA}

\author{Baoyi Chen}
\email{Co-corresponding author: baoyi.chen@tju.edu.cn}
\affiliation{Department of Physics, Tianjin University, Tianjin 300350, China}

\date{\today}
\begin{abstract}

We investigate the production of the $\Omega_{ccc}$ baryon in relativistic heavy-ion collisions. Unlike proton-proton collisions, nuclear collisions produce both deconfined matter and abundant charm quark pairs, which can coalesce to form the $\Omega_{ccc}$ baryon, thereby significantly enhancing its production. We employ the Langevin model and the Instantaneous Coalescence Model (LICM), coupled with hydrodynamic simulations, to study charm quark diffusion and coalescence into the $\Omega_{ccc}$ baryon in expanding QCD matter. The production of the $\Omega_{ccc}$ is governed by the charm quark densities and the in-medium wavefunctions of the $\Omega_{ccc}$, which determines the coalescence probability for the three charm quarks. We calculate the $\Omega_{ccc}$ production with realistic charm diffusions and different in-medium wave functions of $\Omega_{ccc}$ baryon. We find that the production of the $\Omega_{ccc}$ baryon is sensitive to these factors, which aids in understanding its properties in the hot QCD medium.

\end{abstract}
\maketitle


Over the past few decades, extensive studies have suggested that the deconfined matter, also known as Quark-Gluon Plasma (QGP), can be generated in relativistic heavy-ion collisions at the Relativistic Heavy-Ion Collider (RHIC) and the Large Hadron Collider (LHC) through the phase transition~\cite{PHENIX:2004vcz,STAR:2005gfr,Bazavov:2011nk}. Various soft and hard probes have been suggested to study the properties of this hot QCD matter~\cite{Chatterjee:2005de,Song:2010mg,He:2014cla,Qin:2015srf}. About forty years ago, Matsui and Satz proposed the $J/\psi$ meson as a clear signal of the deconfined medium~\cite{Matsui:1986dk}, where its yield is anomalously suppressed due to color screening~\cite{Du:2017qkv} and parton inelastic scattering~\cite{Peskin:1979va,Zhao:2024gxt} in the QGP. The extent of this abnormal suppression in charmonium production reflects the initial energy density of the medium~\cite{Liu:2010ej,Zhao:2020jqu}. However, from RHIC to LHC collision energies, the final production of charmonium is observed to be enhanced~\cite{PHENIX:2006gsi,ALICE:2016flj}, despite the increased energy density of the bulk medium at LHC energies. This enhancement is attributed to the coalescence of charm quarks, a process known as regeneration~\cite{Braun-Munzinger:2000csl,Thews:2000rj,Greco:2003vf,Andronic:2003zv,Yan:2006ve}. The contribution from regeneration becomes more significant when the number of charm quarks is large~\cite{Chen:2017duy}.

Studying the properties of $\Omega_{ccc}$ is crucial to understanding the quark model. Based on Quantum Chromodynamics theory, the production cross section of $\Omega_{ccc}$ is significantly suppressed in proton-proton (pp) collisions, as it requires the formation of three charm-quark-antiquark pairs in a single collision event. However, in nucleus-nucleus collisions, where a large number of charm quarks are generated in the QGP, the production of $\Omega_{ccc}$ is proportional to the cube of the charm quark pair number, i.e., $N_{\Omega_{ccc}}\propto (N_{c\bar c})^3$~\cite{He:2014tga}. Thus, heavy-ion collisions provide a unique environment for searching for $\Omega_{ccc}$ and studying its in-medium properties. 
The coalescence (or regeneration) mechanism has also successfully explained the enhancement of charmonium production and the baryon-to-meson ratio~\cite{Greco:2003xt,Fries:2003vb} in nuclear collisions. 
The hadron production from the coalescence process has also been studied by the Statistical Hadronization Model  (SHM)~\cite{Andronic:2017pug,Andronic:2021erx}, which gives a good description of both light and heavy flavor hadrons. 
Our study will employ the coalescence model to describe the hadronization process of $\Omega_{ccc}$ in heavy-ion collisions.

Heavy quarks are produced in the initial hard scatterings of partons, and their initial momentum distributions are far from equilibrium~\cite{Cacciari:1998it,Cacciari:2001td}. These distributions evolve toward kinetic thermalization due to the strong coupling between the heavy quarks and the bulk medium~\cite{He:2011qa}. This suggests that heavy quarks may become partially kinetically thermalized due to their large mass when they turn into hadrons~\cite{Chen:2017duy}. The momentum distribution of charm quarks directly influences the momentum distribution and yield of $\Omega_{ccc}$. The Langevin model will be used to study the diffusions of heavy quarks.
Another key factor is the probability of coalescence between three charm quarks, which is determined by the Wigner function of the baryon $\Omega_{ccc}$. There is no unified conclusion regarding the wave function of $\Omega_{ccc}$. In addition, hot QCD matter can also modify the wave function of $\Omega_{ccc}$ when produced in QGP due to their sizeable binding energy. These ultimately affect the Wigner function and the production of $\Omega_{ccc}$. In this work, we will consider different wave functions of $\Omega_{ccc}$ to predict the range of $\Omega_{ccc}$ production in nuclear collisions.

Assuming small momentum transfer in each random scattering, the diffusion of heavy quarks is obtained by evolving them event-by-event using the Langevin equation. Elastic collisions and medium-induced radiation of heavy quarks can be incorporated into the Langevin equation~\cite{Cao:2013ita,Yang:2023rgb,Xing:2024qcr,Li:2024wqq},
\begin{align}
    \label{eq-lan}
\frac{d{\bf p}}{dt}=-\eta({\bf p}){\bf p}+{\bf \xi}+\mathbf{{\bf f}_{g}},
\end{align}
Where ${\bf p}$ is the charm quark momentum vector, $\eta(p)$ and ${\bf \xi}$ represent the drag and noise terms, respectively. The drag term is given by $\eta(p) = \kappa / (2 T E_Q)$, where $T$ is the local temperature of the medium and $E_Q = \sqrt{m_Q^2 + p^2}$ is the energy of the heavy quark. The charm quark mass is taken to be $m_c = 1.5$ GeV. The diffusion coefficient in momentum space, $\kappa$, is related to the spatial diffusion coefficient $\mathcal{D}_s$ through the relation $\kappa \mathcal{D}_s = 2T^2$~\cite{Cao:2013ita}. Theoretical model studies and lattice QCD calculations suggest that the value of $\mathcal{D}_s 2 \pi T$ exhibits a weak linear dependence on temperature, which approaches around 2.0 at the critical temperature~\cite{He:2012df,Rapp:2018qla}. Based on studies of the $D$ meson spectrum through Bayesian analysis~\cite{Xu:2017obm} and deep neural networks~\cite{Guo:2023phd}, we adopt effective values for $\mathcal{D}_s 2 \pi T$ of (2.0, 4.0, 6.0) for temperatures above the critical temperature, $T > T_c$, where $T_c$ is taken to be 0.15 GeV. For the noise term, neglecting the momentum dependence, the white noise satisfies the following relation: 
\begin{align}
    \langle \xi^i(t) \xi^j(t')\rangle = \kappa \delta^{ij}\delta(t-t').
\end{align}
Here, \(i, j = (1, 2, 3)\) represents the three spatial dimensions. The medium-induced gluon radiation imparts a recoil force on the heavy quark, \({\bf f}_g = - \frac{d{\bf p}_g}{dt}\). This force can be determined from higher twist calculations~\cite{Zhang:2003wk,Majumder:2009ge} and becomes dominant in the high transverse momentum region. In the low transverse-momentum region, the first and second terms on the right-hand side of the Langevin equation dominate the evolution of the heavy quark's momentum~\cite{Jiang:2022uoe}. The initial positions of heavy quarks are randomly generated based on the density of binary collisions in nuclear collisions. The initial momentum of heavy quarks can be simulated using the FONLL model~\cite{Cacciari:2012ny}. 

During the diffusion process of heavy quarks in QGP, these charm quarks may combine to form a new $\Omega_{ccc}$ via the coalescence process in regions where the local temperature of the medium approaches $T_c$. In these regions, the bound state $\Omega_{ccc}$ after the coalescence process can survive within the medium~\cite{He:2014tga}. The coalescence probability for the three charm quarks is calculated using the wave function of $\Omega_{ccc}$ through the Weyl transform. The wave function of $\Omega_{ccc}$ is denoted as $\psi(r)$, and the Wigner function for the coalescence probability is given by,
\begin{align}
\label{eq-wig}
    W({\bf r},{\bf p})=\int d^6 {\bf y} e^{-i{\bf p}\cdot  {\bf y}} \psi({\bf r}+\frac{\bf y}{2})\psi^\ast ({\bf r}-\frac{\bf y}{2}).
\end{align}
Here, $r$ and $p$ represent the relative distance and relative momentum between the three charm quarks. Since the coalescence process occurs within the deconfined medium, the wave function of $\Omega_{ccc}$ is modified by the hot medium, which in turn alters the Wigner function and affects the final production of $\Omega_{ccc}$. The exact form of the $\Omega_{ccc}$ wave function is not clear, as well as the degree of hot medium modification on the wave function. To account for these factors, we approximate the $\Omega_{ccc}$ wave function as a Gaussian function~\cite{Greco:2003vf}. The width of the Gaussian is varied to reflect the uncertainty in the $\Omega_{ccc}$ wave function and the modifications induced by the hot medium. In the two-body coalescence process, the corresponding Wigner function can be expressed as,
\begin{equation}
W_{M}\left(\boldsymbol{\rho},\boldsymbol{k}_{\rho}\right)=8\exp\left(-\frac{\boldsymbol{\rho}^{2}}{\sigma_{\rho}^{2}}-\boldsymbol{k}_{\rho}^{2}\sigma_{\rho}^{2}\right),
\end{equation}
where the coefficient 8 is a normalization factor. The relative position and momentum are defined in the center-of-mass frame of the two particles,
\begin{align}
\boldsymbol{\rho} & =\frac{1}{\sqrt{2}}\left(\boldsymbol{r}_{1}-\boldsymbol{r}_{2}\right),\\
\boldsymbol{k}_{\rho} & =\sqrt{2}\frac{m_{2}\boldsymbol{k}_{1}-m_{1}\boldsymbol{k}_{2}}{m_{1}+m_{2}},
\end{align}
$\boldsymbol{k}_{i}$, $\boldsymbol{r}_{i}$, and $m_{i}$ represent the momentum, coordinates, and mass of the corresponding particles, respectively. The Gaussian width parameter, $\sigma_{\rho}$, is determined by the mean square radius of the meson, ${\langle r_{M}^{2} \rangle}$, which can be calculated as follows,
\begin{align}
\left\langle r_{M}^{2}\right\rangle  & =\frac{3}{2}\frac{m_{1}^{2}+m_{2}^{2}}{\left(m_{1}+m_{2}\right)^{2}}\sigma_{\rho}^{2}\nonumber \\
 & =\frac{3}{4}\frac{m_{1}^{2}+m_{2}^{2}}{m_{1}m_{2}\left(m_{1}+m_{2}\right)\omega}.
\end{align}
In the second line of the equation, we use $\sigma_{\rho} = 1 / \sqrt{\mu_{1} \omega}$ and $\mu_{1} = 2 \left( 1 / m_{1} + 1 / m_{2} \right)^{-1}$, where $\omega$ is the oscillator frequency~\cite{sun2017analytical}, which can be applied for further calculations of the many-body Wigner function. In the three-body coalescence process, the Wigner function of the baryon is extended as~\cite{wang2019hadronization,Zhao:2020irc}.
\begin{equation}
\label{eq-wig-b}
W_{B}\left(\rho,\lambda,k_{\rho},k_{\lambda}\right)=8^{2}\exp\left(-\frac{\rho^{2}}{\sigma_{\rho}^{2}}-\frac{\lambda^{2}}{\sigma_{\lambda}^{2}}-k_{\rho}^{2}\sigma_{\rho}^{2}-k_{\lambda}^{2}\sigma_{\lambda}^{2}\right)
\end{equation}
where $\boldsymbol{\lambda}$ and $\boldsymbol{k}_{\lambda}$ represent the relative coordinate and relative momentum in the center-of-mass frame for the three quarks that form a baryon.
\begin{align}\boldsymbol{\lambda} & =\sqrt{\frac{2}{3}}\left(\frac{m_{1}\boldsymbol{r}_{1}+m_{2}\boldsymbol{r}_{2}}{m_{1}+m_{2}}-\boldsymbol{r}_{3}\right),\\
\boldsymbol{k}_{\lambda} & =\sqrt{\frac{3}{2}}\frac{m_{3}\left(\boldsymbol{k}_{1}+\boldsymbol{k}_{2}\right)-\left(m_{1}+m_{2}\right)\boldsymbol{k}_{3}}{m_{1}+m_{2}+m_{3}}.
\end{align}
The width parameter $\sigma_{\rho}$ is defined in the same way as in the previous two-body Wigner function. The parameter $\sigma_{\lambda}$ is determined by the relation $\sigma_{\lambda} = 1 / \sqrt{\mu_{2} \omega}$, where the reduced mass is given by $\mu_2 = (3 / 2) \left[ 1 / (m_{1} + m_{2}) + 1 / m_{3} \right]^{-1}$. The value of $\omega$ can be determined using the mean-square radius $\langle r_B^2 \rangle$ of the baryon, as follows,
\begin{align}
\left\langle r_{B}^{2}\right\rangle =\frac{1}{2}\frac{m_{1}^{2}\left(m_{2}+m_{3}\right)+m_{2}^{2}\left(m_{1}+m_{3}\right)+m_{3}^{2}\left(m_{1}+m_{2}\right)}{\left(m_{1}+m_{2}+m_{3}\right)m_{1}m_{3}\omega}.
\end{align}
Here, $\langle r_B^2 \rangle$ is regarded as the mean square radius of the $\Omega_{ccc}$ in the medium at the coalescence temperature, as it will be used to calculate the values of the two Gaussian widths in the Wigner function. The value of $\langle r_B^2 \rangle$ for $\Omega_{ccc}$ is calculated to be $\sqrt{\langle r_\Omega^2 \rangle} = 0.36$ fm~\cite{minissale2024multi}. Other values will also be considered to assess the uncertainty of $\Omega_{ccc}$ wave function on its production. With the spin of the $\Omega_{ccc}$ being 3/2, the statistical factor is $g_B = 1 / 54$, which will be used in the subsequent coalescence formula.

We assume that the coalescence process occurs slightly above the critical temperature, specifically at $T_{\text{coal}} = 1.2\ T_c$, similar to the regeneration of charmonium ground state as their binding energies are close to each other~\cite{Chen:2021akx,He:2014tga}. When charm quarks move to regions with a medium temperature of $T_{\text{coal}}$, their positions and momenta are incorporated into the Wigner function to calculate the coalescence probability. The event-averaged coalescence probability can be expressed as~\cite{Zhao:2020wcd, Chen:2021akx, Han:2016uhh},
\begin{equation}
\begin{array}{l}
\left\langle\mathcal{P}_{{ccc} \rightarrow \Omega}\left(\mathbf{x}_{\mathbf{B}}, \mathbf{p}_{\mathbf{B}}\right)\right\rangle \\
= g_{B} \int  \frac{d \mathbf{x}_{\mathbf{1}}d \mathbf{p}_{\mathbf{1}}}{(2 \pi)^{3}} \frac{d \mathbf{x}_{\mathbf{2}}d \mathbf{p}_{\mathbf{2}}}{(2 \pi)^{3}} \frac{d \mathbf{x}_{\mathbf{3}} d \mathbf{p}_{\mathbf{3}}}{(2 \pi)^{3}} \frac{d^{2} N_{1}}{d \mathbf{x}_{\mathbf{1}} d \mathbf{p}_{\mathbf{1}}} \frac{d^{2} N_{2}}{d \mathbf{x}_{\mathbf{2}} d \mathbf{p}_{\mathbf{2}}} \frac{d^{2} N_{3}}{d \mathbf{x}_{\mathbf{3}} d \mathbf{p}_{\mathbf{3}}}\\
\quad \times W_{B}\left(\rho,\lambda,k_{\rho},k_{\lambda}\right) 
\delta^{(3)}\left(\mathbf{p}_{\mathbf{B}}-\mathbf{p}_{\mathbf{1}}-\mathbf{p}_{\mathbf{2}}-\mathbf{p}_{\mathbf{3}}\right)\\
\quad \times\delta^{(3)}\left(\mathbf{x}_{\mathbf{B}}-\frac{\mathbf{x}_{\mathbf{1}}+\mathbf{x}_{\mathbf{2}}+\mathbf{x}_{\mathbf{3}}}{3}\right).
\end{array}
\end{equation}
Here, $\mathbf{p}_{\mathbf{B}}$ and $\mathbf{x}_{\mathbf{B}}$ represent the momentum and coordinate of the $\Omega_{ccc}$. $dN_i/(d{\bf x}_id{\bf p}_i)$ is the normalized charm density in phase space. The momentum of the $\Omega_{ccc}$ is approximated to be $\mathbf{p}_{\mathbf{B}} = \mathbf{p}_{\mathbf{1}} + \mathbf{p}_{\mathbf{2}} + \mathbf{p}_{\mathbf{3}}$, and the coordinate is expressed as $\mathbf{x}_{\mathbf{B}} = (\mathbf{x}_{\mathbf{1}} + \mathbf{x}_{\mathbf{2}} + \mathbf{x}_{\mathbf{3}})/3$. In Pb-Pb collisions at $\sqrt{s_{NN}} = 5.02$ TeV, multiple charm pairs are produced. The yield of the baryon $\Omega_{ccc}$ in a given rapidity slice $\Delta y$ is given by,
\begin{align}
    N_{B}^{A A}&=\int \frac{d \mathbf{x}_{\mathbf{B}} d \mathbf{p}_{\mathbf{B}}}{(2 \pi)^{3}}\left\langle\mathcal{P}_{{ccc} \rightarrow \Omega}\left(\mathbf{x}_{\mathbf{B}}, \mathbf{p}_{\mathbf{B}}\right)\right\rangle\left(N_{c \bar{c}}^{A A}\right)^{3},\\
    N_{c \bar{c}}^{A A}&=\int d \mathbf{x}_{\mathbf{T}} T_{Pb}(\mathbf{x}_{{T1}}) T_{Pb}(\mathbf{x}_{{T2}}) \mathcal{R}_{S} \frac{d \sigma^{p p}_{c \bar{c}}}{d y} \Delta y,
\end{align}
Here, ${\bf x}_{T1} = {\bf x}_T + {\bf b}/2$ and ${\bf x}_{T2} = {\bf x}_T - {\bf b}/2$. As we simulate the dynamical evolution of three randomly distributed charm quarks in the hot medium, $\left\langle \mathcal{P}_{{ccc} \rightarrow \Omega} \left( \mathbf{x}_{\mathbf{B}}, \mathbf{p}_{\mathbf{B}} \right) \right\rangle$ represents the mean coalescence probability of uncorrelated three charm quarks in the medium. Here, ${\bf b}$ is the impact parameter. $N_{c\bar{c}}^{AA}$ denotes the number of charm pairs produced within the rapidity range $\Delta y$ in Pb-Pb collisions. This number depends on the differential cross section $d\sigma^{pp}_{c\bar{c}}/dy = 1.16$ mb~\cite{ALICE:2021dhb} in the central rapidity region of pp collisions. $T_{Pb} = \int dz \, \rho_{Pb}({\bf x}_T, z)$ is the thickness function, where the nuclear density $\rho_{Pb}$ of Pb is modeled by the Woods-Saxon distribution~\cite{Woods:1954zz, Zheng:2024mep}. The shadowing effect, which modifies nuclear parton densities, also affects the production of charm pairs. The shadowing factor, denoted by $\mathcal{R}_S$, varies with the positions of the parton hard scatterings and changes with the transverse momentum in the reaction. For charm pair production, the shadowing factor averaged over positions in the nucleus is approximately 0.7, as obtained from the EPS09 package~\cite{Eskola:2009uj} in central nuclear collisions, and approaches unity in peripheral collisions.

To determine the dynamical evolution of heavy quarks in the expanding QCD medium, bulk medium properties, such as temperatures and velocities, are required to calculate the drag and noise terms in the Langevin equation. Studies of the light hadron spectrum using hydrodynamic models suggest that the hot QCD matter produced in RHIC and LHC energies behaves as a nearly perfect fluid~\cite{Song:2010mg}. In this work, we use the MUSIC package~\cite{Schenke:2010rr, Schenke:2010nt} to generate the profiles of medium temperatures and velocities. The Equation of State (EoS) for the deconfined phase is taken from lattice QCD results, while the EoS for the hadronic gas is obtained from the Hadron Resonance Gas model~\cite{Huovinen:2009yb}. The two phases are connected by a crossover phase transition. For simplicity, we assume the shear viscosity of the bulk medium is zero. The profiles of the initial medium energy density for $\sqrt{s_{NN}} = 5.02$ TeV Pb-Pb collisions, used as input for the hydrodynamic equations, are generated using the Glauber model.

The production and also properties of $\Omega_{ccc}$ has been studied in relativistic heavy-ion collisions~\cite{He:2014tga,Zhao:2023qww,minissale2024multi}, which indicates a very strong enhancement of $\Omega_{ccc}$ compared to the case in pp collisions. In Ref.[He:2014cla], it
explored the wave functions of the $\Omega_{ccc}$ in vacuum with the potential model,
which are then used to calculate the Wigner function. Meanwhile, charm quarks are usually assumed to achieve kinetic thermalization at
the onset of the hydrodynamic evolution, similar to light partons. This simple approximation allows charm dynamics to be easily described with the conservation equation $\partial_\mu(\rho_c u^\mu)=0$, where $\rho_c$ and $u^\mu$ are the spatial density of the charm and the fluid four-velocity, respectively. However, extensive studies suggest that charm quarks may reach kinetic thermalization with a relaxation time scale. Therefore, in this work, we mainly focus on two aspects: (1) impact of non-thermalized charm quark distributions on the production of $\Omega_{ccc}$, where the realistic momentum of charm quarks is evolved event-by-event by considering different diffusion coefficients $\mathcal{D}_s$. (2) Take different wave functions and the corresponding Wigner function of $\Omega_{ccc}$ used in the coalescence process. The dependence of $\Omega_{ccc}$ production on the dynamics of charm diffusion, as well as the in-medium wave function of $\Omega_{ccc}$ in the coalescence process, will be studied in detail.

\begin{figure}[!hbt]
    \centering
\includegraphics[width=0.45\textwidth]{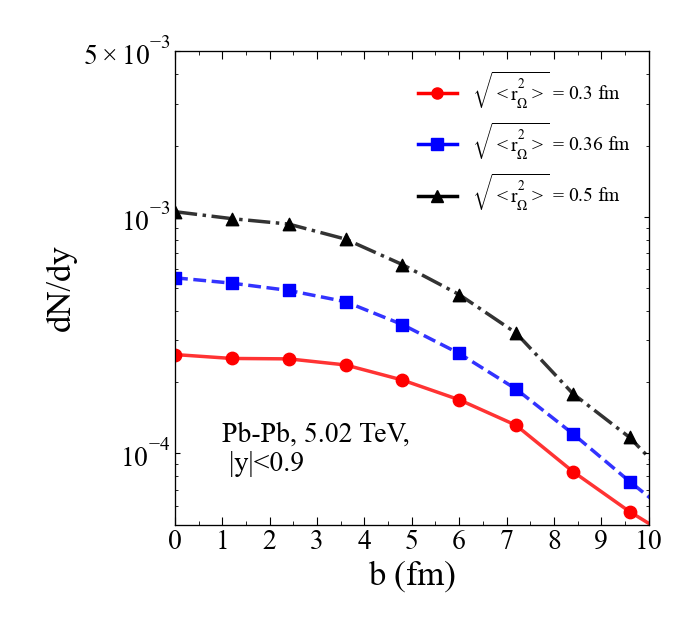}
    \caption{The production of $\Omega_{ccc}$ as a function of the impact parameter in central rapidity $|y| < 0.9$ Pb-Pb collisions at $\sqrt{s_{NN}} = 5.02$ TeV. The spatial diffusion coefficient is taken to be $\mathcal{D}_s 2\pi T = 2.0$. The mean square radius of the $\Omega_{ccc}$ is considered with values of $\sqrt{\langle r_\Omega^2 \rangle}(T_{\rm coal}) = (0.3, 0.36, 0.5)$ fm, in order to explore the dependence of $\Omega_{ccc}$ production on its wavefunction. }
    \label{fig-rB-var}
\end{figure}

In previous work, we employed the Langevin plus Instantaneous Coalescence Model (LICM) to describe the final transverse momentum spectrum of open and hidden charmed mesons in $\sqrt{s_{NN}} = 5.02$ TeV Pb-Pb collisions~\cite{Chen:2021akx}. Our theoretical results are in good agreement with the experimental data for both D and $J/\psi$ mesons. Since $\Omega_{ccc}$ can be thermally produced within hot QCD matter due to its relatively large binding energies~\cite{minissale2024multi,He:2014tga}, the hot medium effects on the wave function of $\Omega_{ccc}$ are encoded in the width of the Wigner function. These medium effects can be approximated by varying the mean square radius, $\langle r_\Omega^2 \rangle$, at $T = T_{\rm coal}$, which in turn determines the values of the widths ($\sigma_\lambda$, $\sigma_\rho$) in the Wigner function. For example, three values of the root-mean-square radius of $\Omega_{ccc}$ are considered in Fig.~\ref{fig-rB-var}, $\sqrt{\langle r_\Omega^2 \rangle}(T_{\rm coal}) = (0.3, 0.36, 0.5)$ fm. The values of 0.36 fm and 0.5 fm are obtained based on Ref.\cite{minissale2024multi} and Ref.\cite{He:2014tga}, respectively. The value of 0.3 represents a scenario with tighter binding for $\Omega_{ccc}$. Charm pair production cross section in pp collisions changes from $d\sigma_{cc}/dy=0.7$ mb to $1.16$ mb when collision energy changes from 2.76 TeV to 5.02 TeV. This is expected to enhance the $\Omega_{ccc}$ production by a factor of $(1.16/0.7)^3\approx 4.5$. In Fig.\ref{fig-rB-var}, our calculation (the line with $\sqrt{\langle r_\Omega^2 \rangle}(T_{\rm coal}) = 0.5$ fm) is around 4.0 times of the $\Omega_{ccc}$ production calculated at 2.76 TeV Pb-Pb collisions given by Ref.\cite{He:2014tga}, where the mean radius of $\Omega_{ccc}$ is around 0.5 fm. The slight suppression of the yield is attributed to a hotter QCD medium generated in Pb-Pb collisions at 5.02 TeV than at 2.76 TeV, which results in a more violent expansion of the bulk medium. This reduces the spatial densities of charm quarks and their coalescence probability~\cite{Zhao:2017yan}. Other types of $\Omega_{ccc}$ wave functions, characterized by $\sqrt{\langle r_\Omega^2 \rangle}(T_{\rm coal}) = (0.3, 0.36)$ fm, are also considered. Tighter wave functions for $\Omega_{ccc}$ impose a stronger constraint on the spatial distance between the triple random charm quarks in the coalescence process. This reduces the yield of $\Omega_{ccc}$ by several times, suggesting that $\Omega_{ccc}$ production in heavy-ion collisions can also aid in exploring the in-medium triple heavy quark potential.

\begin{figure}[!hbt]
    \centering
\includegraphics[width=0.45\textwidth]{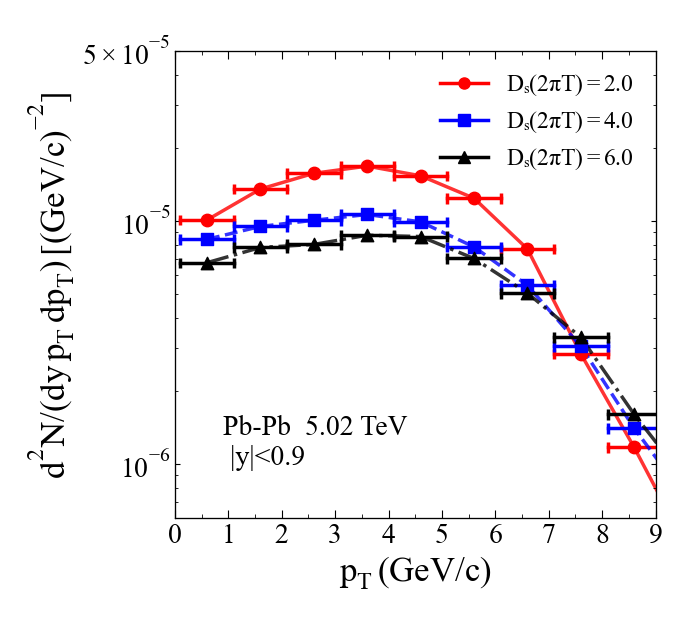}
    \caption{The transverse momentum dependence of $\Omega_{ccc}$ $d^2N/(dyp_T dp_T)$ as a function of $p_T$ for centrality 0-20\% in $\sqrt{s_{NN}} = 5.02$ TeV Pb-Pb collisions. The root mean square radius, which characterizes the wave function of $\Omega_{ccc}$ in the medium, is taken as $\sqrt{\langle r^2_\Omega \rangle}(T_{\rm coal}) = 0.36$ fm. }
    \label{fig-npt-varDs}
\end{figure}

Studies on the collective flows of the charmonium ground state indicate that charm quarks are partially kinetically thermalized during the coalescence of charmonium in the hot QCD medium~\cite{Pan:2023ouw}, despite the fact that $D$ mesons are nearly kinetically thermalized after the kinetic freeze-out. Since the binding energy of $\Omega_{ccc}$ is analogous to that of $J/\psi$, their coalescence temperatures are expected to be similar.
To investigate the effect of different degrees of charm kinetic thermalization on the production of $\Omega_{ccc}$, we consider three values for the spatial diffusion coefficient, as shown in Fig.\ref{fig-npt-varDs}. Smaller values of $\mathcal{D}_s 2\pi T$ indicate a stronger coupling between heavy quarks and the bulk medium, leading to greater kinetic thermalization of charm quarks. As seen in the figure, when charm quarks approach kinetic thermalization, represented by the line for $\mathcal{D}_s 2\pi T = 2.0$, the relative momentum between the triple random charm quarks becomes smaller, which enhances their coalescence probability according to the Wigner function in Eq. (\ref{eq-wig-b}). If charm quarks are less thermalized during the coalescence process, the production of $\Omega_{ccc}$ is significantly reduced, as seen for the line with $\mathcal{D}_s 2\pi T = 6$ in Fig.\ref{fig-npt-varDs}. These three lines define a band for the $\Omega_{ccc}$ production, considering the realistic diffusion of charm quarks in the expanding QCD medium.

\begin{figure}[!hbt]
    \centering
\includegraphics[width=0.45\textwidth]{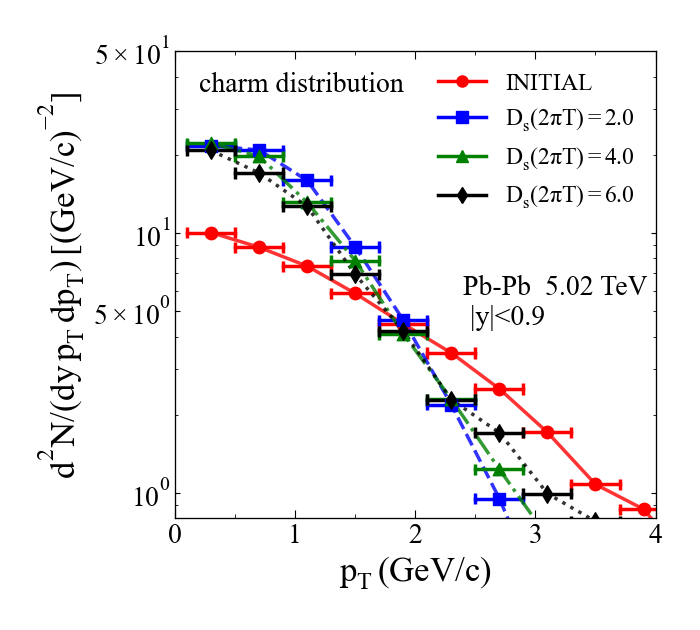}
    \caption{\tred{The transverse momentum distribution of charm quarks  $d^2N/(dyp_T dp_T)$ as a function of $p_T$ for the centrality 0-20\% in $\sqrt{s_{NN}} = 5.02$ TeV Pb-Pb collisions. Four kinds of momentum distributions are plotted: the initial momentum distribution from FONLL calculation, three final momentum distributions with the spatial diffusion coefficient to be $\mathcal{D}_s 2\pi T= 2.0, 4.0, 6.0$ respectively.} }
    \label{fig-npt-c}
\end{figure}

\tred{Since the transverse momentum distribution of charm quarks plays a significant role in determining the transverse momentum distribution of 
$\Omega_{ccc}$ during its production, we have plotted the initial transverse momentum distribution of charm quarks in Fig.\ref{fig-npt-c}, calculated using the FONLL approach. The final-state transverse momentum distributions of charm quarks are obtained using the Langevin equation and plotted in the figure, with varying coupling strengths, $\mathcal{D}_s 2\pi T=2.0, 4.0, 6.0$. The figure shows that the final momentum distribution of charm quarks is softer than the initial distribution. As 
$\Omega_{ccc}$ is produced through the coalescence of three charm quarks, a stronger coupling between the charm quarks and the deconfined medium leads to softer momentum distributions for both the charm quarks and the corresponding 
$\Omega_{ccc}$.
 }
 
In summary, we employ the Langevin equation and the coalescence model to investigate the dynamical evolution of heavy quarks and their coalescence into the triple-charm baryon $\Omega_{ccc}$ in relativistic heavy-ion collisions. The evolution of the bulk medium is simulated using a hydrodynamic model. The binding energy of $\Omega_{ccc}$ remains non-zero even in deconfined matter, indicating that $\Omega_{ccc}$ coalescence occurs within the hot QCD medium. The coalescence probability is linked to the in-medium wave function of $\Omega_{ccc}$. To account for the uncertainty in the $\Omega_{ccc}$ wave function and the modifications induced by the hot QCD medium, we employ different Gaussian functions for the Wigner function. Additionally, the realistic diffusion of charm quarks is considered to study the dependence of $\Omega_{ccc}$ production on the degree of charm kinetic thermalization. These improvements provide more realistic predictions for $\Omega_{ccc}$ production in heavy-ion collisions, ultimately contributing to a deeper understanding of the properties of $\Omega_{ccc}$ at finite temperatures in heavy-ion collisions.

\vspace{3cm}
{\bf Acknowledgments:}
Baoyi Chen is supported by the National Natural Science Foundation of China (NSFC) under Grant No. 12175165.

\bibliography{paper-ccc}


\end{document}